\documentclass{iopart} 

\usepackage{epsf}
\usepackage{latexsym}

\def\submitted{\vspace{28pt plus 10pt minus 18pt}
     \noindent{\small\rm  {\it \journal}\par}}
\def\journal{\ifnum\thejnl=0 \begin{center} Dedicated to Prof. Giorgio Ferrarese in honor of his 70th birthday\end{center} \fi}

\def\beq{\begin{equation}}
\def\eeq{\end{equation}}

\def\xc{\hbox{\rlap{\hskip 1.5pt\raise .75pt\hbox{--}}$\Xscr$}}

\def\lambdabar{\rlap{\hbox{${}^-$}}\lambda}

\def\pmb#1{\setbox0=\hbox{$#1$}%
  \kern-.025em\copy0\kern-\wd0
  \kern.05em\copy0\kern-\wd0
  \kern-.025em\raise.0433em\box0}
\def\pmbs#1{\setbox0=\hbox{$\scriptstyle #1$}%
  \kern-.0175em\copy0\kern-\wd0
  \kern.035em\copy0\kern-\wd0
  \kern-.0175em\raise.0303em\box0}

\def\bfOmega{\pmb{\Omega}}

\def\dualp#1{{}^{\ast_{(\hbox{$\scriptstyle #1$})}} \kern-1pt}

\catcode`@=11  
\def\meqalign#1{\null\,\vcenter{\openup\jot\m@th
  \ialign{\strut\hfil$\displaystyle{##}$&&$\displaystyle{{}##}$\hfil
      \crcr#1\crcr}}\,}
\catcode`@=12  

\def\pmb#1{\setbox0=\hbox{$#1$}%
  \kern-.025em\copy0\kern-\wd0
  \kern.05em\copy0\kern-\wd0
  \kern-.025em\raise.0433em\box0}

\begin{document}

\title[Acceleration-Induced Nonlocality]
{Acceleration-Induced Nonlocality}

\author{
Bahram Mashhoon }

\address{
Department of Physics and Astronomy, University of Missouri-Columbia,\\
Columbia, Missouri 65211, USA
}

\begin{abstract}
The standard relativistic theory of accelerated reference frames in Minkowski spacetime is described.
The measurements of accelerated observers are considered and the limitations of the standard theory, based on the hypothesis of locality, are pointed out. The physical principles of the nonlocal theory of accelerated observers are presented. The implications of the nonlocal theory are briefly discussed.
\end{abstract}


\submitted 

\section{Introduction}

At a fundamental level nonlocality is usually associated with quantum mechanics, since the classical laws of physics are basically local.
This standard description assumes, however, that the primary physical measurements are performed by ideal inertial observers.
On the other hand, all actual observers (including measuring devices, etc.) are noninertial.
It follows that experiments performed with accelerated devices have in fact provided the observational basis for the fundamental laws of physics. This circumstance implies that there must be a connection between real (i.e. noninertial) and inertial observers. The nature of such a relationship is the subject of this paper.

According to the standard theory of relativity, a noninertial observer is at each instant equivalent to an otherwise identical momentarily comoving inertial observer.
This {\it hypothesis of locality} postulates a pointwise equivalence between noninertial and ideal inertial observers. The hypothesis of locality originates from Newtonian mechanics of point particles. The state of a classical particle is determined by its position and velocity at a given instant of time. If the force on the particle is turned off at some instant, the particle will follow the osculating straight line. Thus the assumption of locality is automatically satisfied in this case, since the noninertial and the ideal osculating inertial observer share the same state and are hence equivalent.
This is why the discussion of accelerated systems in classical mechanics does not require any new hypothesis.
In classical electrodynamics,  however, we need to deal with classical electromagnetic waves; their interactions can only be considered poinlike in the geometric optics limit. If all physical phenomena could be reduced to pointlike coincidences of classical particles and rays of radiation, then the hypothesis of locality would be exactly valid \cite{1}.
However, in general classical waves have intrinsic extensions in space and time characterized by their wavelengths and periods. For instance, to measure the  frequency of an incident wave, a few oscillations of the wave must be observed before a reasonable determination of the frequency becomes possible. This situation must be compared with the intrinsic scales of length and time associated with an accelerated observer. That is, an accelerated observer has intrinsic length scales $\mathcal{L}=c^2/a$ or $c/\Omega$  corresponding to its translational acceleration $a$ or rotational frequency $\Omega$ of its spatial frame and the relevant intrinsic time scales are then $c/a$ and $1/\Omega$.
Let $\lambdabar$ be the intrinsic length scale of the phenomenon under observation; then, the hypothesis of locality is valid if $\lambdabar /\mathcal{L} \to 0$, i.e. the deviations from locality characterized by $\lambdabar /\mathcal{L}$ are so small as to be below the sensitivity level of measurements by the accelerated observer \cite{2,3}.
It turns out that this is indeed the case for most situations of interest at present, since for Earth-based devices
$c^2/g \simeq 1 $ lyr and $c/\Omega_{\oplus} \simeq 28$ AU.

A noninertial observer passes through a continuous infinity of hypothetical momentarily comoving inertial observers along its worldline. This is mathematically analogous to the fact that a curved line is the envelope of the infinite class of straight lines tangent to it.
Just as the replacement of a curve by its tangent is only a first approximation, one can show that the hypothesis of locality simply provides an estimate that is exact only in the eikonal limit.
Once the limitations of the hypothesis of locality are recognized, it becomes possible to explore suitable nonlocal alternatives. In this way one arrives at acceleration-induced nonlocality as a  ubiquitous feature of physics even in the classical domain.

It is important to recognize that the hypothesis of locality is nevertheless an integral part of the nonlocal theory described in this paper: in the eikonal limit $\lambdabar /\mathcal{L} \to 0$, the nonlocal theory reduces to the standard theory based on the hypothesis of locality.
This is analogous to the correspondence between wave mechanics and classical mechanics. The idea of such a correspondence will be employed throughout this paper; for instance, we will assume that accelerated observers can perform spatial and temporal measurements that are essentially consistent with the hypothesis of locality (cf. section 2). A nonlocal treatment will be required only if the wave phenomena involved are such that $\lambdabar /\mathcal{L} $ is not negligibly small and hence cannot be ignored.

The standard theory of accelerated systems is described in the following section, where units are chosen such that $c=1$. 
A critique is provided in section 3 and the nonlocal theory of accelerated observers is described.  Section 4 contains a brief discussion of acceleration-induced nonlocality.

\section{Accelerated frames of reference}

Imagine a background global inertial reference frame with coordinates $(t,x,y,z)$ and the class of fundamental observers in this frame. Each fundamental observer is by definition at rest in this frame and carries an orthonormal tetrad frame 
$\tilde \lambda^\mu{}_{(\alpha )}=\delta^\mu{}_{\alpha}$ such that $\tilde \lambda^\mu{}_{(0)}$ is tangent to its worldline and 
$\tilde \lambda^\mu{}_{(i)}$ , $i=1,2,3$, characterize its spatial frame. Consider now an accelerated observer following a worldline $D$ with four-velocity $u^\alpha_D=dx^\alpha_D /d \tau$ and translational acceleration $A^\alpha_D=du^\alpha_D /d \tau$. It is interesting to note that $A_D \cdot u_D =0$ so that $A_D$ is a spacelike vector such that $A_D\cdot A_D=a^2$, where $a$ is the magnitude of the translational acceleration. Here $\tau$ is a temporal parameter along $D$ defined by
$d\tau /d t= \gamma^{-1}$, where $\gamma$ is the Lorentz factor, $\gamma^{-1}=\sqrt{1-v^2(t)}$ and $\mathbf{v}(t)$ is the velocity of the accelerated device. It follows from the hypothesis of locality that the accelerated device is at each instant endowed with an orthonormal tetrad frame $\lambda^\mu{}_{(\alpha )}$ as well such that $\lambda^\mu{}_{(0)}=u^\mu_D$ and
\beq
\eta_{\mu\nu}\lambda^\mu{}_{(\alpha )} \lambda^\nu{}_{(\beta )} =\eta_{\alpha\beta},
\eeq
where $\eta_{\mu\nu}$ is the Minkowski metric tensor with signature $+2$.
Moreover, the application of the hypothesis of locality to the measurement of time by the accelerated observer implies that $\tau$ is the proper time along $D$.
The variation of the orthonormal tetrad along the path of the reference observer is given by
\beq
\label{due}
\frac{d \lambda^\mu{}_{(\alpha )}}{d \tau}=\Phi_\alpha{}^\beta (\tau) \, \lambda^\mu{}_{(\beta )},
\eeq
where $\Phi_{\alpha\beta}$ is an antisymmetric acceleration tensor with \lq\lq electric\rq\rq$\,$  part
$\Phi_{0i}=a_i$ and \lq\lq magnetic\rq\rq$\,$  part $\Phi_{ij}=\epsilon_{ijk}\Omega^k$.
Here $\mathbf{a}$ and $\mathbf{\bfOmega}$ are spacetime scalars that represent respectively the translational acceleration, 
$a_i=A_D \cdot \lambda_{(i )}$, and the rotational frequency of the local spatial frame with respect to the local nonrotating (i.e. Fermi-Walker transported) frame. It is useful to consider the invariants
\beq
I=\frac12 \Phi_{\alpha\beta} \Phi^{\alpha\beta}, \qquad I^*= \frac12 \Phi^*_{\alpha\beta}\Phi^{\alpha\beta},
\eeq
where $\Phi^*_{\alpha\beta}=\frac12 \epsilon_{\alpha\beta\gamma\delta}\Phi^{\gamma\delta}$ is the dual acceleration tensor.
The significance of $I=-a^2+\Omega^2$ and $I^*= -\mathbf{a}\cdot\bfOmega $ lies in the fact that these depend merely on the acceleration, while $\Phi_{\alpha\beta}$ depends  on the velocity as well as the acceleration of the reference observer.
At any given instant of proper time $\tau$, $I$ and $I^*$ are independent  of any  local Lorentz tranformations of the tetrad frame in eq. (\ref{due}); therefore, they represent the velocity-independent  content of the acceleration tensor 
$\Phi_{\alpha\beta}$.
The proper acceleration scales can then be defined in terms of $I$ and $I^*$, i.e. $|I|^{-1/2}$ and $|I^*|^{-1/2}$.

Consider now a geodesic coordinate system established along the worldline $D$ of the fiducial observer.
At any given instant $\tau$ along $D$, the straight spacelike geodesic lines orthogonal to $D$ span a hyperplane that  is  Euclidean space. Let $x^\mu$ be the coordinates of a point on such a hypersurface and let $X^\mu$ be the corresponding geodesic coordinates. Then 
\beq
\label{quattro}
\tau=X^0, \quad x^\mu=x^\mu_D(\tau) +X^i \lambda^\mu{}_{(i )}(\tau)
\eeq
completely characterize the transformation to the new geodesic coordinates. Writing the metric of the background system as
$ds^2=\eta_{\mu\nu}dx^\mu dx^\nu$ and differentiating system (\ref{quattro}), one finds with the help of equation (\ref{due}) that $ds^2=g_{\mu\nu}dX^\mu dX^\nu$, where
\beq
\label{cinque}
g_{00}=-S,\quad g_{0i}=U_i, \quad g_{ij}=\delta_{ij}.
\eeq
Here $S$ and $\mathbf{U}$ are given by 
\beq
\label{sei}
S=(1+\mathbf{a}\cdot \mathbf{X})^2-U^2, \qquad \mathbf{U}=\bfOmega\times \mathbf{X},
\eeq
where $\mathbf{a}$ and $\bfOmega$ are in general functions of $X^0$. One can show that ${\rm det}(g_{\mu\nu})=g$ is given by
\beq
\label{sette}
g=-(1+\mathbf{a}\cdot \mathbf{X})^2
\eeq
so that the inverse metric tensor can be expressed as
\beq
\label{otto}
g^{00}=\frac{1}{g},\quad g^{0i}=-\frac{U^i}{g}, \quad g^{ij}=\delta^{ij}+\frac{1}{g}U^iU^j .
\eeq
A detailed examination of the geodesic coordinate system shows that these coordinates are admissible as long as $g_{00}<0$ \cite{4}. The boundary of the admissible region is characterized by $S=0$. At each instant of time $X^0$, $S=0$ is a quadratic equation in the spatial coordinates  and represents a surface. Such surfaces have been classified under the Euclidean group into seventeen standard forms called quadric surfaces. Specifically,
\beq
\label{nove}
S(X^0,\mathbf{X})=1+2 a_i(X^0) X^i +M_{ij}(X^0)X^iX^j=0,
\eeq
where $M=(M_{ij})$ is a symmetric matrix with components
\beq
\label{dieci}
M_{ij}=a_ia_j+\Omega_i \Omega_j -\Omega^2 \delta_{ij}.
\eeq
It is possible to show that $M$ has eigenvalues
\beq
\label{undici}
\mu_0=-\Omega^2, \qquad \mu_\pm =-I\pm \sqrt{I^2+I^*{}^2},
\eeq
so that $\mu_+ \ge 0$, $\mu_0 \le 0 $, $\mu_- \le 0$ and
\beq
\label{dodici}
{\rm det} (M_{ij})=\mu_+\mu_0\mu_-=\Omega^2 (\mathbf{a}\cdot \bfOmega)^2 .
\eeq

Consider first the general case in which ${\rm det} (M_{ij})\not = 0$. It follows that $M$ has an inverse and it is possible to show that
\beq
\label{tredici}
(M^{-1})_{ij} a^i a^j=1 .
\eeq
The matrix $M$ can be diagonalized at any instant $X^0$ by a rotation of spatial coordinates. The standard form of the quadric surface represented by $S=0$ is then achieved by completing the squares in eq. (\ref{nove}) followed by a translation to new coordinates. More explicitly, let $R$ be the orthogonal matrix such that $R^{-1}M R$ is diagonal with diagonal elements $(\mu_+, \mu_0, \mu_-)$. Using the rotated spatial coordinates $\hat X=R^{-1} X$ and corresponding parameters $\hat a =R^{-1}a$, the translations
\beq
\label{quattordici}
\xi=\hat X_1 +\frac{\hat a_1}{\mu_+}, \quad \eta=\hat X_2 +\frac{\hat a_2}{\mu_0}, \quad 
\zeta=\hat X_3 +\frac{\hat a_3}{\mu_-}, \quad 
\eeq
define a new spatial coordinate system $(\xi , \eta , \zeta )$. In terms of these new coordinates, eq. (\ref{nove})
then takes the form
\beq
\label{quindici}
|\mu_+|\,  \xi^2-|\mu_0|\,  \eta^2-|\mu_-|\,  \zeta^2=0,
\eeq
which represents a {\it real quadric cone} (i.e. an elliptic cone) in general. 
In deriving this result, the relation
\beq
\frac{\hat a_1{}^2}{\mu_+}+\frac{\hat a_2{}^2}{\mu_0}+\frac{\hat a_3{}^2}{\mu_-}=1,
\eeq
which follows from eq. (\ref{tredici}), has been employed. 
An important feature of eq. (\ref{quindici})
should be noted: the extent of validity of the admissible coordinates is determined by the acceleration lengths that are implicit in the eigenvalues of $M$.

If $M$ is a singular matrix, then either $\Omega=0$, in which case the quadric surface degenerates  to {\it coincident planes}, or $\Omega \not= 0$ but $\mathbf{a}\cdot \bfOmega=0$, in which case the quadric surface is a {\it cylinder}.
This cylinder is {\it hyperbolic} for $\Omega^2 < a^2$ and {\it parabolic} for $\Omega^2=a^2$. It is a real {\it elliptic}
cylinder for $\Omega^2 > a^2$. These assertions can be simply demonstrated by working in a system of coordinates that is obtained from the $(X_1,X_2,X_3)$ system by a rotation such that in the new system one coordinate axis is parallel to $\mathbf{a}$  and another is parallel to $\bfOmega$. For $a=0$, eq. (\ref{nove}) reduces to a circular cylinder of radius $\Omega^{-1}$.

It is instructive to consider for instance an observer rotating uniformly with frequency $\Omega_0$ about the $z-$axis on a circle of radius $r$ in the $(x,y)-$plane. The natural orthonormal tetrad of the observer is given by
\begin{eqnarray}
\lambda^\mu{}_{(0 )}&=& \gamma (1,-v\sin \phi, v\cos \phi,0), \nonumber \\
\lambda^\mu{}_{(1 )}&=& (0,\cos \phi, \sin \phi,0), \nonumber \\
\lambda^\mu{}_{(2 )}&=& \gamma (v,-\sin \phi, \cos \phi,0), \nonumber \\
\lambda^\mu{}_{(3 )}&=& (0,0, 0,1) ,
\end{eqnarray}
where $\phi=\Omega_0 t =\gamma \Omega_0 \tau$, $v=r\Omega_0$ and $\phi=0$ at $t=\tau=0$ by assumption.
It follows from eq. (\ref{due}) that the only nonzero components of $\mathbf{a}$ and $\bfOmega$ are the centripetal acceleration $a_1=-v \gamma^2 \Omega_0$ and the rotation frequency relative to ideal gyroscope directions $\Omega_3=\gamma^2 \Omega_0$, so that $\mathbf{a}\cdot \bfOmega=0$ in this case and $\Omega^2 > a^2$.
The invariants of the acceleration tensor are $I=\gamma^2 \Omega_0^2$ and $I^*=0$;
hence $\mathcal{L}=1/(\gamma \Omega_0)$ is the proper acceleration length of the observer. The geodesic coordinate system
$(T,X,Y,Z)$ is related to the inertial coordinate system $(t,x,y,z)$ via
\begin{eqnarray}
t&=&\gamma (T+vY), \nonumber \\
x&=&(X+r)\cos(\gamma \Omega_0 T) - \gamma Y \sin (\gamma \Omega_0 T), \nonumber \\
y&=&(X+r)\sin(\gamma \Omega_0 T) + \gamma Y \cos (\gamma \Omega_0 T), \nonumber \\
z&=&Z .
\end{eqnarray}
The reference observer is at the spatial origin of the new coordinates, which are admissible within the surface
\beq
(X+r)^2+\gamma^2 Y^2 =\frac{1}{\Omega_0^2}.
\eeq
This elliptic cylinder for any constant $Z$ has semimajor axis $\Omega_0^{-1}$, semiminor axis $\Omega_0^{-1} \sqrt{1-v^2}$
and eccentricity $v$. For $Z=0$, the center of the ellipse is at the origin of the background inertial system and the reference observer is at one of the foci of this ellipse; as $v\to 1$, the reference observer approaches the so-called light cylinder at $r=\Omega_0^{-1}$ and the area of the ellipse tends to zero.

In general, the boundary hypersurface $S(X)=0$ can be timelike, spacelike or null. To see this, consider the scalar quantity
$\mathcal{N}= N \cdot N$, where $N_\mu = \frac12 \partial S/\partial X^\mu$ is normal to $S=0$.
Using eqs. (\ref{sette}) and (\ref{otto}), $\mathcal{N}$ can be written as 
\beq
\label{ventuno}
\mathcal{N} =- W^2 +2W(\mathbf{a}\times \bfOmega)\cdot \mathbf{X}+ [\bfOmega + (\mathbf{a}\cdot \bfOmega)\mathbf{X}]^2,
\eeq
where $W$ is given by
\beq
\label{ventidue}
W = \dot \mathbf{a}\cdot \mathbf{X}-\frac{
(\dot{ \bfOmega} \times \mathbf{X})\cdot (\bfOmega \times \mathbf{X})
}{
1+\mathbf{a}\cdot \mathbf{X}
}
\eeq
and an overdot indicates differentiation with respect to $X^0$. In the absence of rotation ($\Omega=0, \dot{\mathbf{a}}\not =0$), the hypersurface $S=0$ is spacelike; however, it becomes null for uniform translational acceleration for all time, which is unphysical due to the requirement that an external source must supply an infinite amount of energy in this case. 
To avoid such unphysical situations, the acceleration should in general be turned on at a finite time $\tau_0$ and then turned off at some later time. In the absence of translational acceleration ($\mathbf{a}=0$), the hypersurface is timelike for the case of uniform rotation ($\dot{\bfOmega}=0$). This conclusion is consistent with the fact that while observers on the Earth generally use the Earth-based rotating coordinate system, the corresponding boundary hypersurface (i.e. light cylinder) does not hinder any outside radiation  from reaching the Earth and vice versa.

\section{Nonlocality}

It is necessary to confront the geometric description of the accelerated reference system presented in the previous section with the measurements of noninertial observers on the basis of the hypothesis of locality. The question is whether the spacelike geodesic segments away from the fiducial worldline and all the way to the boundary are in fact measurable.
A detailed examination of the issue of length measurement in accelerated systems has revealed that the hypothesis of locality provides a unique answer only when the length involved is negligibly small compared to the relevant acceleration length of the observer \cite{2,3,5}. This means that one is in effect confined to the immediate neighborhood  of the worldline of the reference observer. Noninertial coordinate systems --- such as the Earth-based systems used in everyday life --- are definitely useful. As a matter of principle, however, the extent of validity of such systems is rather limited \cite{2,3,5}.
Similar restrictions are expected to hold for other possible ways of constructing accelerated coordinate systems \cite{6}.

Acceleration-induced nonlocality originates from the conflict between the hypothesis of locality and wave measurements of accelerated observers. Consider, for instance, the measurement of the frequency of a plane monochromatic wave incident on an observer rotating uniformly with frequency $\Omega_0$ about the $z-$axis. Replacing the accelerated observer at each instant by a hypothetical momentarily comoving inertial observer in accordance with the hypothesis of locality, one can connect the resulting local inertial frames with the background global inertial system by Lorentz transformations. Such transformations can be used in two ways. The first method uses the invariance of the phase of the wave and results in the Doppler 
effect \cite{7}
\beq
\label{ventitre}
\omega_D'=\gamma (\omega - \mathbf{v}\cdot \mathbf{k}),
\eeq
where $\omega$ and $\mathbf{k}$ are respectively the frequency and propagation vector of the plane wave as determined by the fundamental inertial observers. That is,  $\omega_D'=-u_D^\mu k_\mu$, where $k^\mu=(\omega, \mathbf{k})$. The second method is based on a pointwise determination of the electromagnetic field of the wave by the accelerated observer,
\beq
\label{ventiquattro}
F_{(\alpha )(\beta )}= F_{\mu\nu} \lambda^\mu{}_{(\alpha )}\lambda^\nu{}_{(\beta )},
\eeq
which is then Fourier analyzed in terms of the observer's proper time $\tau$ with the result that
\beq
\label{venticinque}
\omega' =\gamma (\omega -m \Omega_0),
\eeq 
where $m=0,\pm 1, \pm 2, \ldots $. Here $\hbar m$ has the interpretation of the $z-$component  of the total angular momentum of the radiation field \cite{7}. In general $\omega_D'$ is time-dependent, while $\omega'$ represents a constant spectrum. The reception of a few oscillations of the wave would be necessary in order to determine its frequency;
therefore, it follows that the instantaneous result $\omega_D'$ must be valid in the eikonal limit $\lambdabar /\mathcal{L} \to 0$. To see how this could come  about  using the general formula (\ref{venticinque}), consider the eikonal approximation, where the electromagnetic radiation can be represented by a ray. The total angular momentum of the radiation field can then be expressed as $\mathbf{j}=\mathbf{r}\times \mathbf{p} + \mathbf{s}$, where $\mathbf{p}=\hbar \mathbf{k}$ and $\mathbf{s}$
is the spin vector. Thus in this approximation eq. (\ref{venticinque}) takes the form
\beq
\hbar \omega' \simeq \gamma (\hbar \omega - \mathbf{j}\cdot \bfOmega_0)\simeq \hbar \gamma (\omega -\mathbf{v}\cdot \mathbf{k}) - \gamma \mathbf{s} \cdot \bfOmega_0,
\label{ventisette}
\eeq
where $\mathbf{v}=\bfOmega_0\times \mathbf{r}$ is the velocity of the observer. Here the term $- \gamma \mathbf{s} \cdot \bfOmega_0$ indicates the phenomenon of spin-rotation coupling \cite{7}, which vanishes in the eikonal limit $\lambdabar /\mathcal{L} \to 0$ and hence $\omega'$ reduces to $\omega_D'$ in this limit. 

The general result (\ref{venticinque}) has, however, a drawback in comparison with $\omega_D'$: $\omega'$  can be negative or zero.
That $\omega'$ can be negative does not lead to any basic difficulty, since the noninertial character of the rotating observer is absolute \cite{8}. On the other hand, $\omega'=0$ means that for $\omega=m\Omega_0$ the radiation can be made to stand completely still by a mere rotation of the observer. This contradicts the spirit of relativity theory: electromagnetic radiation can never stand completely still with respect to any inertial observer.

As emphasized by Bohr and Rosenfeld, the measurement of the electromagnetic field involves an averaging process over a region of spacetime \cite{9,10}. For the noninertial reference observer, the averaging must be done in the accelerated reference frame; however, only the past worldline of the observer need be taken into account due to the uniqueness problem of length
measurement in accelerated frames \cite{2,3,5}. Let $\mathcal{F}_{(\alpha)(\beta)}(\tau )$ be the actual field measured by  
the accelerated observer along its worldline and $F_{(\alpha)(\beta)}(\tau )$ be the field measured by the instantaneously comoving inertial observer. The most general linear relationship between $\mathcal{F}_{(\alpha)(\beta)}$ and $F_{(\alpha)(\beta)}$ consistent with causality is \cite{11}
\beq
\label{ventisette}
\mathcal{F}_{(\alpha)(\beta)}(\tau )= F_{(\alpha)(\beta)}(\tau )+ \int_{\tau_0}^\tau K_{\alpha\beta}{}^{\gamma\delta}(\tau, \tau ') F_{(\gamma)(\delta)}(\tau ')d \tau',
\eeq
where $\tau_0$ is the instant at which the acceleration is turned on and $K_{\alpha\beta\gamma\delta}$ is a kernel that is expected to be proportional to the acceleration
of the observer. For a radiation field with $\lambdabar /\mathcal{L} \to 0$, 
the nonlocal part of the ansatz (\ref{ventisette}) is expected to vanish. The nonlocal ansatz 
(\ref{ventisette}) deals only with spacetime scalars and is thus manifestly 
invariant under inhomogeneous Lorentz transformations of the background spacetime.

A significant feature of the Volterra integral relation (\ref{ventisette}) is that the relationship between
$\mathcal{F}_{(\alpha)(\beta)}$ and $F_{(\alpha)(\beta)}$ is unique in the space of continuous functions according to Volterra's theorem \cite{12}.
This important result has been extended to the Hilbert space of square-integrable functions by Tricomi \cite{13}.
The connection between ${F}_{(\alpha)(\beta)}$ and $F^{\mu\nu}$ along the worldline is unique as well by eq. (\ref{ventiquattro}); therefore, the main uniqueness relation extends to $\mathcal{F}_{(\alpha)(\beta)}$ and $F^{\mu\nu}$. Expressing eq. (\ref{ventiquattro}) in matrix notation such that its right side has the form $\Lambda F$, 
where $\Lambda$ is a $6\times 6$ matrix and $F$ is a column vector with electric and magnetic field components $(\mathbf{E},\mathbf{B})$, one can write
ansatz (\ref{ventisette}) in matrix form as
\beq
\label{ventotto}
\mathcal{F}(\tau)= \Lambda (\tau) F(\tau )+ \int_{\tau_0}^{\tau} K(\tau, \tau ') \Lambda (\tau') F(\tau ')d\tau' ,
\eeq
where $\mathcal{F}(\tau_0)= \Lambda (\tau_0) F(\tau_0 )$ and it remains to determine the kernel $K$. The unique connection between $\mathcal{F}$ and $F$ can be used to exclude the circumstance that a basic radiation field could stand completely still with respect to a noninertial observer. To this end, one can postulate that if $\mathcal{F}$ is a constant, then $F$ must be a constant as well, i.e.
\beq
\label{ventinove}
\Lambda (\tau_0) =\Lambda (\tau)+ \int_{\tau_0}^{\tau} K(\tau, \tau ') \Lambda (\tau')d\tau',
\eeq
which simply follows from equation (\ref{ventotto}) in this case. This Volterra integral equation can be used to determine $K$. With such a kernel, any true radiation field $F$ --- which is not a constant by definition --- would lead to a variable $\mathcal{F}$
by the Volterra-Tricomi uniqueness theorem. Hence the nonlocal theory is so formulated that a basic radiation field will never stand completely still with respect to any observer. This approach is a direct generalization of the situation regarding inertial observers: an inertial observer moving along the direction of propagation of an electromagnetic wave measures a frequency $\omega_D'=\gamma\omega (1-v)$ and $\omega'_D=0$ if and only if $\omega=0$; that is, if one inertial observer measures a constant field, it must be constant for all inertial observers.

A detailed examination of the solutions of eq. (\ref{ventinove}) has revealed that there is indeed a unique kernel that leads to finite results and is consistent with all available observational data \cite{14}. This solution is given by 
\beq
K(\tau,\tau')=\kappa (\tau '), 
\eeq
where the {\it kinetic kernel} $\kappa$ can be expressed as
\beq
\label{30}
\kappa (u) = - \frac{d \Lambda(u)}{du}\Lambda^{-1}(u) .
\eeq
In this way, one arrives at a unique nonlocal theory of accelerated observers \cite{14,15}.

Electromagnetic field measurements have been considered thus far for the sake of simplicity and comparison with observational data. However,  the nonlocal theory is generally applicable to any basic radiation field. 
In this connection, an immediate consequence of this theory should be noted: acceleration-induced nonlocality
rules out the existence  of fundamental scalar (or pseudoscalar) fields. Such a field would have to be local according to equation (\ref{30}); then, eq. (\ref{venticinque}) would imply that the scalar radiation could stand completely still with respect to a rotating observer in violation of our basic postulate. The nonlocal theory thus predicts that scalar (or pseudoscalar) fields should be composites. The absence of fundamental scalar (or pseudoscalar) fields in nature is in agreement with observational data available at present.

\section{Discussion}

It is important to mention a direct consequence of nonlocality in the case of helicity-rotation coupling. Imagine an observer rotating with constant frequency $\Omega_0$ in the positive sense about  the direction of propagation of a plane wave of definite helicity and frequency $\omega\gg \Omega_0$. It turns out that eq. (\ref{venticinque}) is valid according to the nonlocal theory except when $\omega'=0$; in this case ($\omega=m\Omega_0$), the field exhibits resonance behavior.
For the situation under consideration here eq. (\ref{venticinque}) reduces to $\omega' =\gamma (\omega \mp \Omega_0)$, where the upper (lower) sign refers to an incident positive (negative) helicity wave. Moreover, it follows from the nonlocal ansatz (\ref{ventisette}) that the amplitude of the positive (negative) helicity component is enhanced (diminished) by a factor $1+\Omega_0/\omega$ ($1-\Omega_0/\omega$) to first order in $\Omega_0/\omega \ll 1$.
It would be of great interest to subject this prediction of the theory to experimental test.

Should the train of thought presented in this work turn out to be fruitful, then it would seem rather likely that gravitation even at the classical level would have to be described by a nonlocal field theory in Minkowski spacetime as a consequence of Einstein's principle of equivalence. The development of a nonlocal field theory of gravitation that in the eikonal limit would have an interpretation in terms of spacetime curvature, as in general relativity, remains a task for the future.

\section*{Acknowledgments}
I am grateful to D. Bini and C. Chicone for helpful discussions.


\section*{References}

\begin{thebibliography}{00}

\bibitem{1}
Einstein A 1950 
{\it The Meaning of Relativity}, Princeton University Press, Princeton

\bibitem{2}
Mashhoon B 1990
{\it Phys. Lett.} {\bf A143} 176

\bibitem{3}
Mashhoon B 1990
{\it Phys. Lett.} {\bf A145} 147

\bibitem{4}
Lichnerowicz A 1955
{\it Th\'eories Relativistes de la Gravitation  et de l'\'Electromagn\'etisme},
Masson, Paris

\bibitem{5}
Mashhoon B and Muench U 2002
{\it Ann. Phys. (Leipzig)} {\bf 11} 532

\bibitem{6}
Marzlin K-P 1996
{\it Phys. Lett.} {\bf A215} 1


\bibitem{7}
Mashhoon B 2002
{\it Phys. Lett.} {\bf A306} 66

\bibitem{8}
Mashhoon B 1986
{\it Found. Phys. }{\bf 16} (Wheeler Festschrift) 619 

\bibitem{9}
Bohr N and Rosenfeld L 1933
{\it Det. Kgl. dansk. Vid. Selskab.} {\bf 12} n. 8. Translated in: {Quantum Theory and Measurement},
edited by J.A. Wheeler and W.H. Zurek, Princeton University Press, Princeton, 1983

\bibitem{10}
Bohr N and Rosenfeld L 1950
{\it Phys. Rev.} {\bf 78} 794

\bibitem{11}
Mashhoon B 1993
{\it Phys. Rev.} {\bf A47} 4498

\bibitem{12}
Volterra V 1959
{\it Theory of Functionals and of Integral and Integro-Differential Equations}, 
Dover, New York

\bibitem{13}
Tricomi F G 1957
{\it Integral Equations}, 
Interscience, New York

\bibitem{14}
Chicone C and Mashhoon B 2002
{\it Phys. Lett.} {\bf A298} 229


\bibitem{15}
Hehl F W and Obukhov Y N 2003
{\it Foundations of Classical Electrodynamics}, 
Birkh\"auser,  Boston

\end{thebibliography}
\end{document}